\documentclass[amsmath,amssymb,prl,twocolumn,showpacs]{revtex4}
\usepackage[dvips]{graphicx}
\usepackage{bm}

\begin{document}
\title{Bidirectional Single-Electron Counting and the Fluctuation Theorem}

\author{
Y. Utsumi$^1$,
D. S. Golubev$^2$, 
M. Marthaler$^{3}$, 
K. Saito$^4$, 
T. Fujisawa$^{5,6}$, 
and 
Gerd Sch\"on$^{2,3}$
}

\affiliation{
$^1$Institute for Solid State Physics, University of Tokyo, Kashiwa, Chiba 277-8581, Japan \\
$^2$ Forschungszentrum Karlsruhe, Institut f\"ur Nanotechnologie, 76021 Karlsruhe, Germany \\
$^3$Institut f\"{u}r Theoretische Festk\"{o}rperphysik and DFG Center for Functional Nanostructures (CFN), Universit\"{a}t Karlsruhe, 76128 Karlsruhe, Germany \\
$^4$Graduate School of Science, University of Tokyo, Tokyo 113-0033, Japan \\
$^5$NTT Basic Research Laboratories, NTT Corporation, 
Morinosato-Wakamiya, Atsugi 243-0198, Japan \\
$^6$Research Center for Low Temperature Physics, Tokyo Institute of Technology, Ookayama, Meguro, Tokyo 152-8551, Japan 
}

\begin{abstract}
We investigate theoretically and experimentally the full counting statistics of bidirectional single-electron tunneling through a  double quantum dot in a GaAs/GaAlAs heterostructure and compare with predictions of the fluctuation theorem (FT) for Markovian stochastic processes. 
We observe that the quantum point contact electrometer used to study the transport
induces nonequilibrium shot noise and dot-level fluctuations and strongly modifies the tunneling statistics. 
As a result, the FT appears to be violated. 
We show that it is satisfied if the back-action of the electrometer is taken into account, and we provide a quantitative estimate of this effect.
\end{abstract}

\date{\today}
\maketitle

\newcommand{\mat}[1]{\mbox{\boldmath$#1$}}
\newcommand{\mtau}{\mbox{\boldmath$\tau$}}
\newcommand{\cgf}{{\cal W}}

According to the second law of thermodynamics, the entropy of a macroscopic system driven out of equilibrium increases with time until equilibrium is reached. Thus the dynamics of such a system is irreversible. 
In contrast, for a mesoscopic system performing a random trajectory in phase space and measured during a sufficiently short time, the entropy may either increase or decrease. 
The `Fluctuation Theorem' (FT), 
which relies only on the microreversiblity of the underlying equation of motion, states that the probability distribution $P_\tau(\Delta S)$ for processes increasing or decreasing the entropy by $\pm \Delta S$ during a time interval $\tau$ obeys the relation
\begin{eqnarray}
P_\tau(\Delta S)/P_\tau(-\Delta S)=\exp(\Delta S).
\label{DS}
\end{eqnarray}
Remarkably, the FT remains valid even far from equilibrium. 
It has been proven for thermostated Hamiltonian systems~\cite{Evans}, 
Markovian stochastic processes~\cite{Lebowitz,Andrieux}, and 
 mesoscopic conductors~\cite{Tobiska,FB,SU,Esposito,Andrieux1}. 
The FT is fundamentally important for transport theory, 
one of its consequences being the Jarzynski equality~\cite{Jarzynski,Campisi},
which, in turn, leads to the 2nd law of thermodynamics. 
It also leads to the fluctuation-dissipation theorem and Onsager symmetry relations \cite{Gallavotti}, as well as to their extensions to nonlinear transport~\cite{Tobiska,FB,SU,Esposito,Andrieux1}. 

In electron transport experiments the entropy production
is related to Joule heating, $\Delta S= qeV_S/T$,
where $q$ is the number of electrons (with charge  $e$) transfered through the conductor during time $\tau$, and $T$ is the temperature.
Hence the FT can be formulated in terms of the distribution of transfered charge  
$P_\tau(q)$ at sufficiently long times, $\tau\gtrsim e/I$, as follows
\begin{eqnarray}
P_\tau(q)/P_\tau(-q)=\exp\left(qeV_{\rm S}/T\right). 
\label{pft}
\end{eqnarray} 

The FT has been experimentally verified in chemical physics at room temperature~\cite{Wang}, 
while tests of the FT in mesoscopic transport experiments in milli-Kelvin regime have been lacking so far. 
On the other hand, recent advances in time-resolved charge detection by a 
quantum point contact (QPC)~\cite{Fujisawa,Gustavsson,Gustavsson2,Gasser} made it possible 
to measure the distribution $P_\tau(q)$ for single-electron tunneling through quantum dots. 
This opens the possibility of  testing the FT in mesoscopic transport. 

In this article we report on experiments on the direction-resolved full counting statistics (FCS) of single-electron tunneling in a double quantum dot (DQD) system, which is probed by an asymmetrically coupled QPC~\cite{Fujisawa}. 
We analyze the experimental data in the frame of the FT and find that the form (\ref{pft}) appears to be violated. 
However, it is still satisfied, if we replace the temperature by an enhanced value $T^*$. 
We attribute the apparent overheating to the back action of the QPC detector and provide quantitative estimates of the effect. 
Moreover, we note that the form (\ref{DS}) of the FT, valid for Markovian stochastic processes defined by general rates~\cite{Lebowitz,Andrieux}, is satisfied, and we relate the entropy change to a ratio of the relevant nonequilibrium tunneling rates.

\begin{figure}
\includegraphics[width=0.9 \columnwidth]{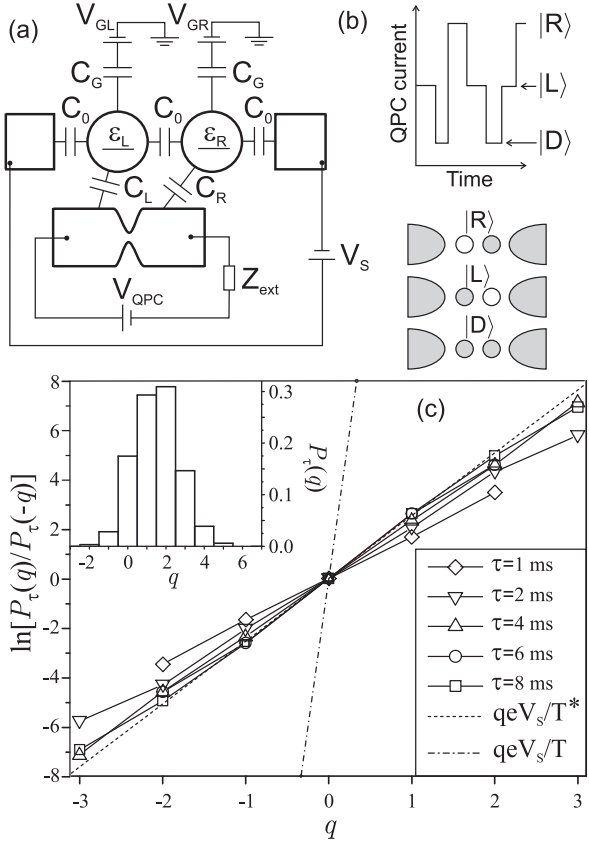} 
\caption{ (a) Schematics of the system.
Only one energy level in each dot, with the energies $\varepsilon_L$ and $\varepsilon_R$ respectively, 
contributes to the transport. 
(b) The QPC current switches between three values corresponding to the three
charge states. 
(c) Test of FT (\ref{pft}) at several times. Lines with symbols: logarithm of lhs of Eq. (\ref{pft}) 
at several times; 
dashed line: 
$q\, eV_S/T^*$  with $T^*=1.37$ K; 
dot-dashed line: $q\, eV_S/T$. 
Inset: the distribution $P_\tau(q)$ at $\tau=4$ ms. 
} 
\label{fig:1}
\end{figure}

{\it Experimental test of the FT} --
The setup of our experiment~\cite{Fujisawa} is shown in Fig.~\ref{fig:1}. 
It consists of the DQD coupled to the QPC detector (Fig.~\ref{fig:1}a). 
The left and right gate voltages $V_{\rm GL},V_{\rm GR}$ applied to the quantum dots are tuned in such a way that only 
three charge states of the DQD are allowed: 
$|L\rangle$, $|R\rangle$, and $|D\rangle$, 
denoting states where the left or right dot is occupied with a single electron,  
or where both dots are occupied, respectively. 
Accordingly, the current through the QPC, which is coupled asymmetrically to the DQD, switches between three different values (Fig.~\ref{fig:1}b). 
This setup allows distinguishing electron tunneling in different directions. 
For example, the switching  $|L\rangle \to | R \rangle$ corresponds to the transfer of one electron from the left dot to the right one, 
while $|R \rangle \to |L \rangle$ signals a transfer in  opposite direction. 
The time trace of the current taken over a sufficiently long time $\tau$ allows one to determine the distribution of transfered charges $P_\tau(q)$.
An example of such a distribution is shown in the inset of Fig.~\ref{fig:1}c.

First, we perform a direct test of the FT, see Fig.~\ref{fig:1}c.
The combination $\ln[P_\tau(q)/P_\tau(-q)]$ indeed is a linear
function of the transfered charge $q$. 
However, the slope  approaches at long time $\tau$ the value $eV_S/T^*$ where
$V_{\rm S} = 300$ $\mu$V is the value of the applied DQD bias voltage, but the effective temperature $T^*=1.37$ K is a fit parameter (dashed line), 
which strongly exceeds the bath temperature of the leads of $T=130$ mK (dot-dashed line). 

\begin{figure}
\includegraphics[width=0.9 \columnwidth]{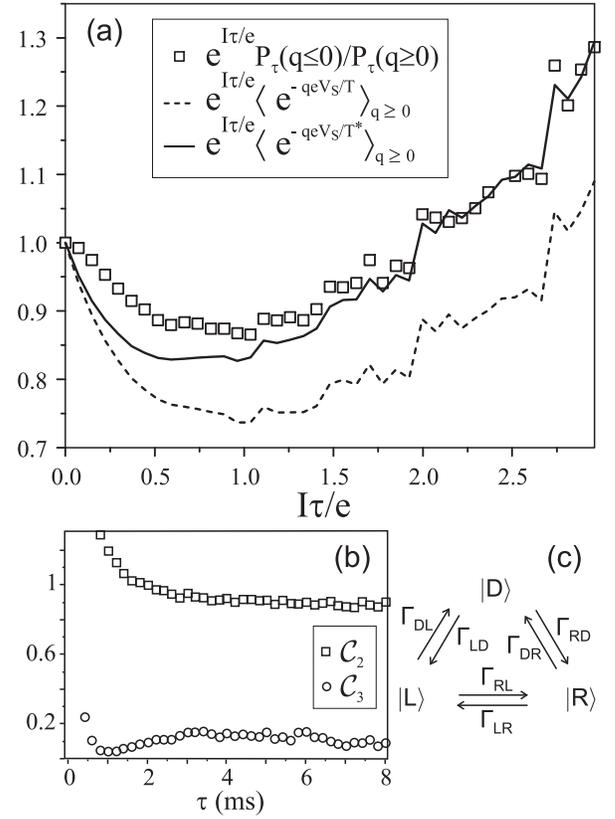} 
\caption{
(a) Test of the integrated FT (\ref{eq:ift}) in the time interval 
$0 \!<\! \tau \!<\! 8$ ms.
The time $\tau$ is multiplied by the frequency of electron tunneling 
$I/e \! \approx \! 370$ Hz. 
Squares: lhs of Eq. (\ref{eq:ift}) (denoted as $P_\tau(q  \leq  0)/P_\tau(q  \geq  0)$);
dashed line: rhs of Eq. (\ref{eq:ift}) (denoted as $\langle \exp(-q eV_{\rm S}/T) \rangle_{q \geq 0}$); 
solid line:  rhs of Eq. (\ref{eq:ift}) with T replaced by $T^* = 1.37$ K. 
All three curves are multiplied by $\exp(I\tau/e)$ for clarity.
(b) Normalized second, 
${\cal C}_2=\langle (q-\langle q\rangle)^2 \rangle/\langle q\rangle$, 
and third cumulants, 
${\cal C}_3=\langle (q-\langle q\rangle)^3 \rangle/\langle q\rangle$, 
of the charge distribution $P_\tau(q)$. 
(c) The six transitions with $\Gamma_{ij}$ between three charge states.
}
\label{fig:2}
\end{figure}

To further test the time dependence contained in Eq. (\ref{pft}), we check the integrated form of the FT (Fig. \ref{fig:2}a), 
\begin{eqnarray}
\frac{\sum_{q=-\infty}^0 P_\tau(q)}{\sum_{q=0}^{\infty} P_\tau(q)}
=\frac{\sum_{q=0}^{\infty} P_\tau(q) \,{\rm e}^{-q eV_{\rm S}/T}}
{\sum_{q=0}^{\infty} P_\tau(q)}.
\label{eq:ift}
\end{eqnarray}
Plotting both sides of Eq.~(\ref{eq:ift}) with the value of the electron temperature, $T=130$ mK, we observe a clear discrepancy between both 
(open squires and dashed line). 
However, by adjusting the temperature in the right hand side of Eq.~(\ref{eq:ift}) to the effective temperature (solid line), we get a good fit for 
$\tau\gtrsim e/I$. 
As we will discuss below, the apparent heating is caused by the back-action of the QPC electrometer, which operates in highly nonequilibrium conditions. 

The experimental distribution of the transfered charge (inset of Fig.~\ref{fig:1}c) deviates strongly  from a Gaussian shape. 
For illustration the second and the third cumulants of the charge distribution are plotted in Fig.~\ref{fig:2}b.
The normalized third cumulant remains close to 0.1 at all times.
Since the FT (\ref{pft}) is satisfied for any Gaussian distribution $P_\tau(q)$, the existence of higher cumulants is a further indication of non-trivial behavior.

{\it Violation and recovery of the FT}--
Why does the FT appear to be violated? 
Quite generally, the FT for the system with four leads, depicted in Fig.~\ref{fig:1}a, should be formulated in terms of the joint probability distribution $P_\tau(q,q')$ of two charges $eq$ and $eq'$ transmitted through the DQD and the QPC, respectively~\cite{SU}, 
\begin{eqnarray}
P_\tau(q,q') = 
\exp \left[ 
\big(q e V_{\rm S}+q' e V_{\rm QPC}\big)/T \right] P_\tau(-q,-q'), 
\label{P2}
\end{eqnarray}
where $T$ is the temperature of the leads. 
Since only the charge $e q$ is measured, Eq.~(\ref{P2}) should be integrated over $e q'$ and afterwards, the right hand side reduces to $\exp[eV_{\rm S}/T]\,
P_\tau(-q)$ only if $V_{\rm QPC}\!=\!0$, which leads to the apparent violation of the FT. 

Then why is the FT recovered by introducing the effective temperature $T^*$? 
In the experiment, the DQD is in the sequential tunneling regime. 
Then the probability distribution $P_\tau(q)$ can be derived from a master equation approach of the FCS~\cite{Andrieux, Bagrets,Belzig}. 
It begins with the `modified' master equation
\begin{eqnarray}
\partial_t \, {\bm p}(t) = {\bm \Gamma}(\lambda) \, {\bm p}(t),
\;\;\;\;
{\bm p}^T=(p_L,p_R,p_D) \, ,
\label{mmastereq}
\end{eqnarray}
where $p_s$ is the occupation probabilities of the charge states ($s=L,R,D$). 
The transition matrix has a modified form ${\bm \Gamma}(\lambda)$ depending on the counting field $\lambda$ measuring the electron transfer through the barrier between the two quantum dots
\begin{eqnarray}
{\bm \Gamma}(\lambda)
\!=\!
\left[ 
\begin{array}{ccc}
-\Gamma_{ RL} - \Gamma_{ DL} & \Gamma_{ LR} \, {\rm e}^{-i\lambda} & \Gamma_{ LD} \\
\Gamma_{ RL} \, {\rm e}^{i\lambda} &  -\Gamma_{ LR} - \Gamma_{DR} & \Gamma_{ RD} \\
\Gamma_{ DL} & \Gamma_{ DR} & -\Gamma_{LD} - \Gamma_{RD}
\end{array}
\right].~
\end{eqnarray}
The characteristic function (CF) takes the form~\cite{Bagrets}
\begin{eqnarray}
{\cal Z}_\tau(\lambda)
&=
\sum_q P_\tau(q) {\rm e}^{iq\lambda}
=
{\bm e}^T 
{\rm e}^{\tau {\bf \Gamma} (\lambda)}
{\bm p}^{st},
\label{Z1}
\end{eqnarray}
where 
${\bm e}^T = (1,1,1)$ and 
${\bm p}^{st}$ is the stationary state, which is found from the equation 
${\bf \Gamma} (0) \, {\bm p}^{st} \!=\! 0$. 
In the long-time limit, 
$\tau \! \gg \! e/I$, 
the CF acquires an exponential form 
${\cal Z}_\tau(\lambda) \approx {\rm e}^{\tau {\cal F}(\lambda)}$,
where 
${\cal F}(\lambda)$ 
is the eigenvalue of the matrix 
${\bf \Gamma}(\lambda)$ 
with the largest real part. 
It satisfies 
\begin{eqnarray}
0=\det[ {\bf \Gamma}(\lambda) - {\cal F} \, {\bf I}]
 = {\cal F}^3 + K {\cal F}^2 + K' {\cal F} \hspace{1.5cm}
\nonumber\\
-\Gamma_{ DR}\Gamma_{ RL}\Gamma_{ LD}({\rm e}^{i\lambda} - 1)
- \Gamma_{ DL} \Gamma_{ LR} \Gamma_{ RD}({\rm e}^{-i\lambda} - 1),
\label{eigen}
\end{eqnarray}
where ${\bf I} $ is the identity matrix. 
Since the parameters $K$ and $K'$ are independent of $\lambda$, we observe
without solving Eq. (\ref{eigen}), that the CF satisfies the identity,
\begin{eqnarray} 
{\cal Z}(\lambda) 
\! &=& \!
{\cal Z}(-\lambda + ieV_{\rm S}/T^*),
\label{FTF} 
\\
T^* 
\! &=& \!
eV_{\rm S}/\ln\left[(\Gamma_{DR}\Gamma_{RL}\Gamma_{LD})/(\Gamma_{DL}\Gamma_{LR}\Gamma_{RD})\right].~
\label{teff}
\end{eqnarray}
Performing the inverse Fourier transform of Eq.~(\ref{FTF}), we arrive at the FT in the form (\ref{pft}) with $T$ replaced by $T^*$. 
The argument of the logarithm in Eq.~(\ref{teff}) is the ratio of products of the tunneling rates corresponding to forward and backward cycles (i.e., the counterclockwise and clockwise cycles shown in Fig.~\ref{fig:2}c). 
The logarithm of this ratio gives the entropy production associated with the transfer of one electron through the system~\cite{Lebowitz,Andrieux}. 
While in the present example we find this transparent result we note that in more general systems with many transport cycles, it is in general not possible to define a unique effective temperature~\cite{Andrieux}. 

In the experiments the tunneling rates are estimated as 
$\Gamma_{DR} \!=\! 4 {\rm kHz}$, 
$\Gamma_{RD} \!=\! 0.3 {\rm kHz}$, 
$\Gamma_{DL} \!=\! 1 {\rm kHz}$, 
$\Gamma_{LD} \!=\! 1.5 {\rm kHz}$, 
$\Gamma_{LR} \!=\! 1.7 {\rm kHz}$, 
and 
$\Gamma_{RL} \!=\! 1.8 {\rm kHz}$. 
The values lie in a regime where they do not suffer from the finite band width of the QPC detector ($\sim \! 10$ kHz). 
The effective temperature derived from Eq.~(\ref{teff}), $T^* \!=\! 1.14$ K,
 agrees well with the value $1.37$ K obtained directly from the FT. 

{\it QPC back-action} --
The tunnel rates are influenced by environmental effects. 
In particular, the DQD is necessarily coupled to the QPC, as modeled by capacitors $C_L$ and $C_R$ in Fig.~\ref{fig:1}a. 
The nonequilibrium current noise of the QPC~\cite{Thielmann} produces fluctuations of the potentials of the quantum dots $e\delta V_{L,R}$, which in turn influence the tunnel rates~\cite{Gustavsson2,Aguado}
as known from the so-called $P(E)$-theory~\cite{Ingold}.
Introducing three such functions $P_j(E)$ with $j=L,R,dd$ we find
\begin{eqnarray}
\Gamma_{RL}  &=&  2 \pi \, |t_{dd}|^2  P_{dd}(E_L-E_R), 
\label{GammaRL} \\
\Gamma_{DL}  &=&  \Gamma_{R} \! \int \! d E \, f(E_D-E_L-\mu_R-E) P_R(E),
\label{GammaDL} \\
\Gamma_{DR}  &=&  \Gamma_{L} \! \int \! d E \, f(E_D-E_R-\mu_L+E)  P_L(E). 
\label{GammaDR}
\end{eqnarray}
Here $t_{dd}$ is the matrix element describing  tunneling between the two dots, the rates 
$\Gamma_L$ and $\Gamma_R$ characterize the coupling  between the
 dots and the leads. 
$f(E)$ is the Fermi function, where we fix the chemical potentials of the leads as $\mu_L \!=\! -\mu_R \!=\! eV_{\rm S}/2$. 
The energies of the charge states $E_{L,R,D}$ include the electrostatic energy of the electric field stored in the capacitors. 
The rates $\Gamma_{LR}$, $\Gamma_{LD}$ and $\Gamma_{RD}$ are given 
by the same expressions where the Fermi function $f$ should be replaced by 
$1\!-\!f$ 
and the argument of the functions $P_j$ should be taken with  opposite sign.

The spectral functions $P_{L/R/dd}$ are expressed in terms of the phase operators 
$\hat\varphi_{L/R}(t) \!=\! \int^t d t' \delta \hat V_{L/R}(t')$ 
and 
$\hat\varphi_{dd} \!=\! \hat\varphi_{R} \!-\! \hat\varphi_{L}$ 
as follows
\begin{eqnarray}
P_j(E)  = 
\int  \frac{dt}{2\pi}
{e}^{i E t}
\left \langle 
{e}^{ i \hat{\varphi}_j(t)}{e}^{-i \hat{\varphi}_j(0)} 
\right \rangle\, 
\approx 
\int  \frac{dt}{2\pi}
{e}^{i E t}
\nonumber \\ 
\times
\exp\left[
\int  \frac{d \omega}{2\pi} 
\frac{e^2[ S_{0,j}(\omega)+S^{\rm QPC}_{V, j} (\omega)]
( {e}^{-i \omega t}- 1)}{\omega^2} 
\right].
\label{ppc}
\end{eqnarray}
Here $S_{0,j}(\omega)$ is the spectral density of thermal fluctuations of equilibrium environments, including the impedance of the external circuit $Z_{\rm ext}$ (Fig.~\ref{fig:1}a), etc. 
The  part $S^{\rm QPC}_{V, j}$ within the Gaussian approximation is proportional to the non-equilibrium and non-symmetrized current noise of the QPC as 
$S^{\rm QPC}_{V,j} (\omega)
=  \kappa_j^2 |Z_t(\omega)|^2 S^{\rm QPC}_I(\omega)$~\cite{Aguado}, 
where 
\begin{eqnarray}
S_I^{\rm QPC}
= \frac{e^2}{\pi}
\left[ \sum_\pm \frac{{\cal T}(1 - {\cal T})
(\omega \pm eV_{\rm QPC})}{1-{\rm e}^{-(\omega \pm eV_{\rm QPC})/T}}
+ \frac{2 {{\cal T}}^2  \omega}{1-{\rm e}^{-\omega/T}}\right].
\nonumber
\end{eqnarray}
Here ${\cal T}$ is the QPC transparency and 
$Z_t$ is the impedance of the electromagnetic environment seen by the QPC; 
$Z_t(\omega) = 1/(-i \omega  \bar{C} + 1/\bar{R})$ 
where the resistance and  capacitance are 
$\bar{R} \!=\! 1/(R_{\rm QPC}^{-1} \!+\! Z_{\rm ext}^{-1})$ 
and 
$\bar{C} \!=\! (3 C_0 \!+\! C_G)/2 \!+\! C_LC_R/(C_L \!+\! C_R)$. 
The factors $\kappa_j$ characterize the coupling between QPC and quantum dots given by certain ratios of the capacitances. 

When the QPC is in equilibrium $V_{\rm QPC} \!=\! 0$, the functions $P_j$ obey the detailed balance 
$
P_j(E)/P_j(-E)={\rm e}^{E/T}
$
and thus 
$\Gamma_{\! RD}/\Gamma_{\! DR}
\!=\!
{\rm e}^{(E_D-E_R-\mu_L)/T}$, etc. 
From Eqs.~(\ref{teff}-\ref{GammaDR}) it follows that in this case $T^* \!=\! T$.
However, in the experiment, the QPC is biased at a rather high voltage 
$eV_{\rm QPC} \! \gg \! T$ 
and the tunnel rates through the central barrier are approximately 
$
\Gamma_{RL/LR} \! \approx \! \Gamma_{\max}/[1+(E_R-E_L)^2/\Gamma_0^2]
$, 
where 
$\Gamma_0= {\cal T}(1-{\cal T}) (e^2\kappa_{dd}\bar R)^2 eV_{\rm QPC}/2\pi$ 
and 
$\Gamma_{\max}  \!=\! 2 \, |t_{dd}|^2/\Gamma_0$. 
From the experimental values $\Gamma_{\max} \approx 7$ kHz, $\Gamma_0\approx 30\mu$eV, $V_{\rm QPC} \!=\! 800$ $\mu$V and ${\cal T} \!=\! 0.19$, 
the parameters can be roughly estimated as $|t_{dd}|\approx 30$ MHz, $\kappa_{dd}\bar R \approx 5$ k$\Omega$. 
The latter value is in agreement with other experiments, see i.e.~\cite{Gustavsson2}. 
Since the difference between the tunnel rates in  opposite directions is significantly reduced, $\Gamma_{LR} \! \approx \! \Gamma_{RL}$, 
we observe that the effective temperature $T^*$ is enhanced. 
From Eq.~(\ref{teff}) we roughly estimate 
$T^* \! \sim \! eV_S/ \ln[\pi eV_S /\Gamma_0]\approx 1$ K, in good agreement with our previous findings, thus further supporting the QPC back-action model.
In order to reduce $T^*$ one can reduce $V_{\rm QPC}$ or the external impedance seen by the QPC $\bar R$.

Other environmental effects do not change our findings by much. 
In GaAs nanostructures acoustic phonons modify the tunneling properties via piezoelectric and deformation coupling. 
At experimental values of the QPC current $I_{\rm QPC} \approx 12 {\rm nA}$ the phonons stay almost in equilibrium~\cite{Gasser}. 
Therefore, the phonon effect is absorbed in the equilibrium part of the spectral density $S_{0,j}$, and does not affect the FT. 
An {\it intrinsic} back-action of the QPC is often addressed in the context of quantum measurement~\cite{Averin}. 
However, in the present case we estimate this effect to be negligible. 
Detailed discussions of these environmental effects and 
 derivations of the theory presented in this article, based on the real-time diagrammatic technique~\cite{Schoeller,Thielmann}, will be published elsewhere~\cite{inpr}.

{\it Summary} --
We have investigated experimentally and theoretically the fluctuation theorem for Markovian stochastic processes by studying the direction-resolved full counting statistics in a double quantum dot system via a nearby quantum point contact electrometer. 
We found an apparent violation of the fluctuation theorem, which we attribute to the non-equilibrium electromagnetic fluctuations generated by the shot noise of the quantum point-contact electrometer. 
We also demonstrated that the FT is recovered if we adopt an effective value for the temperature ten times higher than the electron temperature. 
This effective temperature depends on the entropy production associated with the transfer of one electron through the double quantum dot, which in turn can be expressed by a ratio of forward to backward tunneling rates. 

We would like to thank  M. Hettler for  helpful discussions.
This work has been supported by Strategic International Cooperative Program 
the Japan Science and Technology Agency (JST) and 
by the German Science Foundation (DFG).

\end{document}